\documentclass[aps,pra,showpacs,groupedaddress,twocolumn]{revtex4}
\usepackage{amsfonts}
\usepackage{graphicx}
\usepackage{latexsym}
\usepackage{makeidx}
\usepackage{amsmath}
\usepackage{amssymb}
\usepackage{graphicx}

\begin{document}

\title[MENDART in cavity QED under Lindbladian dephasing]{Mean excitation numbers
due to anti-rotating term (MENDART) in cavity QED under Lindbladian dephasing}
\author{A. V. Dodonov}
\affiliation{Instituto de F\'{\i}sica, Universidade de Bras\'{\i}lia, PO Box 04455,
70910-900, Bras\'{\i}lia, Distrito Federal, Brazil}

\begin{abstract}
We study the photon generation from arbitrary initial state in cavity QED
due to the combined action of the anti-rotating term present in the Rabi
Hamiltonian and Lindblad-type dephasing. We obtain a simple set of
differential equations describing this process and deduce useful formulae
for the moments of the photon number operator, demonstrating analytically that
the average photon number increases linearly with time in the asymptotic limit.
\end{abstract}


\maketitle

In the 2008-th paper by \textit{Werlang et al.} \cite{Werlang} a puzzling
quantum effect was noticed from numerical simulations: when a two-level atom
interacts with a single mode of the radiation field in a cavity by means of
the Rabi Hamiltonian, while subject to standard Markovian dephasing
mechanism, the average intracavity photon number exhibits a linear
growth with time. Such asymptotic photon
generation due to decoherence occurs because for pure dephasing processes
the environment may be viewed as a unmonitored detector making random
nondemolition measurements of the number of quanta in the atom-field system
\cite{Carm,WM}, whilst in \cite{PRL,Nature,Zeno} it was shown that
nondemolition measurements can pump energy into the system via the
destruction of quantum coherence provided the anti-rotating term is kept in
the light-matter interaction Hamiltonian (i.e., without performing the Rotating Wave Approximation \cite{JC}).
Besides, the pure dephasing reservoirs always possess a finite temperature
(see, e.g. \cite{Carm,WM} for microscopic deduction) and hence store an
infinite amount of energy, so the additional system energy is continuously supplied by
the environment and the First Law of Thermodynamics is not violated (for the
discussion concerning the Second Law of Thermodynamics in systems subject to
frequent quantum measurements see \cite{Nature}).

Although the phenomenon of photon generation due to decoherence was
explained qualitatively in \cite{Werlang,CAMOP}, no satisfactory
analysis was carried out to derive analytically whether for the pure Markovian dephasing the average
photon number de facto increases linearly with time and whether this growth saturates for large times. So the aim of this
paper is to investigate analytically the behavior of Mean Excitation Numbers
due to Anti-Rotating Term (MENDART), such as mean photon number and its
variance or atomic excitation probability, and investigate their asymptotic
characteristics in the simplest case of Markovian dephasing. We shall show
that for any initial state in the asymptotic limit the mean photon number $%
\left\langle n\right\rangle $ indeed increases linearly with time, the
average value of the photon number second moment $\left\langle
n^{2}\right\rangle $ grows quadratically with time, and the atomic
excitation probability $P_{e}$ attains a constant value. So this paper
provides the missing mathematical explanation for the phenomenon of steady
photon generation due to Lindblad-type decoherence in the presence of the anti-rotating term.

Our starting point is the Markovian master equation for the density matrix $%
\rho $ that takes into account both the atomic and cavity field
phase-damping (dephasing) \cite{Carm,WM,CAMOP,amendart}%
\begin{equation}
\dot{\rho}=-i[H,\rho ]+\frac{\gamma _{a}}{2}(\sigma _{z}\rho \sigma
_{z}-\rho )+\gamma _{c}\left( 2n\rho n-n^{2}\rho -\rho n^{2}\right) \, \label{111},
\end{equation}%
where $\gamma _{a}$ ($\gamma _{c}$) is the atomic (cavity) dephasing rate
and $H$ is the Rabi Hamiltonian \cite{Rabi,Ra1} (we set $\hbar =1$)%
\begin{equation}
H=\omega n+\frac{\Omega }{2}\sigma _{z}+g(a+a^{\dagger })(\sigma _{+}+\sigma
_{-})  \label{Rabi}
\end{equation}%
that includes the anti-rotating term $(a\sigma _{-}+a^{\dagger }\sigma _{+})$%
. Here $a$ and $a^{\dagger }$ are the cavity annihilation and creation
operators, $n\equiv a^{\dagger }a$ is the photon number operator, and $%
\omega $, $\Omega $ and $g$ are the cavity frequency, the atomic transition
frequency and the atom-field coupling constant, respectively. The Pauli
operators are defined as $\sigma _{-}=|g\rangle \langle e|$, $\sigma
_{+}=|e\rangle \langle g|$ and $\sigma _{z}=|e\rangle \langle e|-|g\rangle
\langle g|$, so that kets $|g\rangle $ and $|e\rangle $ can be interpreted
as atomic\ ground and excited states, respectively.

Expanding the density matrix in the Fock basis as%
\begin{eqnarray}
\rho  &=&\sum_{n,m=0}^{\infty }(a_{n,m}|g,n\rangle \langle
g,m|+b_{n,m}|e,n\rangle \langle e,m|  \nonumber \\
&&+c_{n,m}|g,n\rangle \langle e,m|+c_{m,n}^{\ast }|e,n\rangle \langle g,m|)~,
\end{eqnarray}%
where $a_{n,m}$, $b_{n,m}$ and $c_{n,m}$ are time-dependent coefficients, we
obtain the exact set coupled differential equations (the prime stands for
the time derivative)%
\begin{eqnarray}
a_{n,m}^{\prime } &=&i[\omega (m-n)+i\gamma _{c}\left( n-m\right)
^{2}]a_{n,m}+ig(\sqrt{m}c_{n,m-1}  \nonumber \\
&-&\sqrt{n}c_{m,n-1}^{\ast }+\sqrt{m+1}c_{n,m+1}-\sqrt{n+1}c_{m,n+1}^{\ast })
\label{a1}
\end{eqnarray}%
\begin{eqnarray}
b_{n,m}^{\prime } &=&i[\omega (m-n)+i\gamma _{c}\left( n-m\right)
^{2}]b_{n,m}+ig(\sqrt{m}c_{m-1,n}^{\ast }  \nonumber \\
&-&\sqrt{n}c_{n-1,m}+\sqrt{m+1}c_{m+1,n}^{\ast }-\sqrt{n+1}c_{n+1,m})
\label{a2}
\end{eqnarray}%
\begin{eqnarray}
c_{n,m}^{\prime } &=&if_{n,m}c_{n,m}+ig(\sqrt{m}a_{n,m-1}  \label{a3} \\
&&-\sqrt{n+1}b_{n+1,m}+\sqrt{m+1}a_{n,m+1}-\sqrt{n}b_{n-1,m}),  \nonumber
\end{eqnarray}%
where $f_{n,m}\equiv \omega \left( m-n\right) +\Omega +i[\gamma _{a}+\gamma
_{c}(n-m)^{2}]$. In the strong dephasing limit, $(\gamma _{a}+\gamma _{c})\gtrsim |g|$, one
expects on physical ground that due to the decoherence the terms $c_{n,m}$
rapidly attain some constant values, so assuming that $c_{n,m}^{\prime }=0$
we get%
\begin{eqnarray}
c_{n,m} &=&\frac{g}{f_{n,m}}(\sqrt{n+1}b_{n+1,m}-\sqrt{m}a_{n,m-1}  \nonumber
\\
&&+\sqrt{n}b_{n-1,m}-\sqrt{m+1}a_{n,m+1})\,.  \label{cc}
\end{eqnarray}%
Now we substitute the expression for $c_{n,m}$ back into equations (\ref{a1}%
)--(\ref{a2})\footnote{%
Actually, one must have in mind that the equation (\ref{cc}) only holds
after sufficient amount of time, but such nuances are not relevant when one
is interested in the asymptotic behavior.} and define new coefficients $%
\tilde{a}_{n,m}=e^{-i\omega t(m-n)}a_{n,m}$ and $\tilde{b}_{n,m}=e^{-i\omega
t(m-n)}b_{n,m}$ that are slowly varying functions of time. Assuming that $%
\left\vert g\right\vert \ll \omega ,\Omega $ (a condition that holds in cavity QED experiments unless the
so-called `ultra-strong coupling regime' \cite{Hows1,Bla} is achieved) we neglect the rapidly
oscillating terms and obtain the following effective differential equations
for the diagonal probability amplitudes%
\begin{equation}
\tilde{a}_{n}^{\prime } =-[(v_{1}+v_{2})n+v_{2}]\tilde{a}%
_{n}+[v_{1}nb_{n-1}+v_{2}(n+1)\tilde{b}_{n+1}]  \label{d1}
\end{equation}
\begin{equation}
\tilde{b}_{n}^{\prime } =-[(v_{1}+v_{2})n+v_{1}]\tilde{b}%
_{n}+[v_{2}na_{n-1}+v_{1}( n+1) \tilde{a}_{n+1}],  \label{d2}
\end{equation}
where $\tilde{a}_{n}\equiv \tilde{a}_{n,n}$, $\tilde{b}_{n}\equiv \tilde{b}%
_{n,n}\ $and we defined coefficients%
\begin{equation}
v_{1}=\frac{2\gamma g^{2}}{\left( \omega -\Omega \right) ^{2}+\gamma ^{2}}%
,~v_{2}=\frac{2\gamma g^{2}}{\left( \omega +\Omega \right) ^{2}+\gamma ^{2}}~
\end{equation}%
with $\gamma \equiv \gamma _{c}+\gamma _{a}$ standing for the total
dephasing rate.

One can easily verify that the normalization condition is maintained, $%
\sum_{n=0}^{\infty }(\tilde{a}_{n}^{\prime }+\tilde{b}_{n}^{\prime })=0$, so
the equations (\ref{d1})-(\ref{d2}) are consistent and lead to the following
coupled differential equations for the low-order MENDART%
\begin{equation}
\left\langle n(t)\right\rangle ^{\prime }=v_{2}+\left( v_{1}-v_{2}\right)
\left[ P_{e}(t)+\left\langle n\sigma _{z}(t)\right\rangle \right]  \label{n}
\end{equation}%
\begin{equation}
P_{e}(t)^{\prime }=v_{2}-\left( v_{1}+v_{2}\right) \left[ P_{e}(t)+\left%
\langle n\sigma _{z}(t)\right\rangle \right]  \label{Pe}
\end{equation}%
\begin{eqnarray}
\left\langle n\sigma _{z}(t)\right\rangle ^{\prime } &=&v_{2}-2\left(
v_{1}-v_{2}\right) \left\langle n(t)\right\rangle  \label{ns} \\
&&-\left( v_{1}+v_{2}\right) \left[ P_{e}(t)+\left\langle n\sigma
_{z}(t)\right\rangle +2\left\langle n^{2}\sigma _{z}(t)\right\rangle \right]
\nonumber
\end{eqnarray}%
\begin{eqnarray}
\left\langle n^{2}(t)\right\rangle ^{\prime } &=&2\left( \omega ^{2}+\Omega
^{2}+\gamma ^{2}\right) ^{-1}  \label{n2} \\
&&\times \left[ \gamma g^{2}\left( 1+4\left\langle n(t)\right\rangle \right)
-\omega \Omega \left\langle n\sigma _{z}(t)\right\rangle ^{\prime }\,\right]
.  \nonumber
\end{eqnarray}%
These equations cannot be solved analytically due to the coupling to the
dynamical variable $\left\langle n^{2}\sigma _{z}(t)\right\rangle $ which
obeys another differential equation.

However, one can deduce the \textit{general} formula for the average photon
number $\left\langle n(t)\right\rangle $ by noticing the similarity in the
last terms of equations (\ref{n}) and (\ref{Pe}). One gets
\begin{equation}
\left\langle n(t)\right\rangle -\left\langle n(0)\right\rangle =\frac{2}{%
\omega ^{2}+\Omega ^{2}+\gamma ^{2}}\left\{ g^{2}\gamma t-\omega \Omega
\lbrack P_{e}(t)-P_{e}(0)]\right\}\,,  \label{nn}
\end{equation}%
where $P_{e}(t)\leq 1$ is still an unknown function of time\footnote{%
Notice that the obtained asymptotic photon generation rate $2\gamma
g^{2}(\omega ^{2}+\Omega ^{2}+\gamma ^{2})^{-1}$ resembles the approximate
formula obtained in \cite{We} [namely $2\gamma g^{2}((\omega +\Omega
)^{2}+\gamma ^{2})^{-1}$], although in that paper the mathematical approach
was oversimplified.}. Furthermore, in the asymptotic regime $t\rightarrow
\infty $ we expect from the equation (\ref{Pe}) that $P_{e}$ attains a
constant value. Imposing $\lim_{t\rightarrow \infty }P_{e}(t)^{\prime }=0$
we obtain from (\ref{Pe})--(\ref{ns})%
\begin{equation}
\lim_{t\rightarrow \infty }[P_{e}(t)+\left\langle n\sigma
_{z}(t)\right\rangle ]=\frac{1}{2}-\frac{\omega \Omega }{\omega ^{2}+\Omega
^{2}+\gamma ^{2}}  \label{x1}
\end{equation}%
\begin{equation}
\lim_{t\rightarrow \infty }\left\langle n\sigma _{z}(t)\right\rangle
^{\prime }=0\mathrm{\,}
\end{equation}%
\begin{equation}
\lim_{t\rightarrow \infty }\left\langle n^{2}\sigma _{z}(t)\right\rangle =-%
\frac{2\omega \Omega }{\omega ^{2}+\Omega ^{2}+\gamma ^{2}}%
\lim_{t\rightarrow \infty }\left\langle n(t)\right\rangle
\end{equation}%
and from equation (\ref{n2}) we get%
\begin{equation}
\lim_{t\rightarrow \infty }\left\langle n^{2}(t)\right\rangle ^{\prime }=%
\frac{2\gamma g^{2}}{\omega ^{2}+\Omega ^{2}+\gamma ^{2}}\left[
1+4\lim_{t\rightarrow \infty }\left\langle n(t)\right\rangle \right] ~.
\label{x4}
\end{equation}

Hence, in the asymptotic regime $t\rightarrow \infty $ we have the following
rules for the Asymptotic MENDART (AMENDART, as coined in \cite{amendart})
for any initial state: \textbf{a)} $\left\langle n(t)\right\rangle $ and $%
-\left\langle n^{2}\sigma _{z}(t)\right\rangle $ increase \emph{linearly}
with time; \textbf{b)} $\left\langle n^{2}(t)\right\rangle $ increases \emph{%
quadratically} with time; \textbf{c)} $P_{e}(t)$ and $\left\langle n\sigma
_{z}(t)\right\rangle $ attain \emph{constant} values. We solved numerically
the effective differential equations (\ref{d1})--(\ref{d2}) and verified
that the formula (\ref{nn}) is correct for all times, thereby accounting for
the linear growth of $\left\langle n(t)\right\rangle $ noticed in \cite%
{Werlang} from numerical data, while the equations (\ref{x1})-(\ref{x4})
agree with the numerical results in the asymptotic regime.

\begin{figure}[th]
\begin{center}
\includegraphics[width=.49\textwidth]{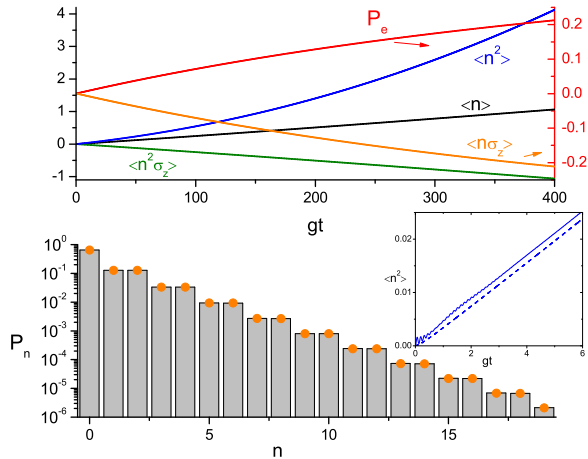}
\end{center}
\caption{Exact and effective dynamical behavior of principal observables and
the photon statistics for the time $gt=300$ in the resonant regime, $\Omega =%
\protect\omega $.}
\label{fig1}
\end{figure}

\begin{figure}[th]
\begin{center}
\includegraphics[width=.49\textwidth]{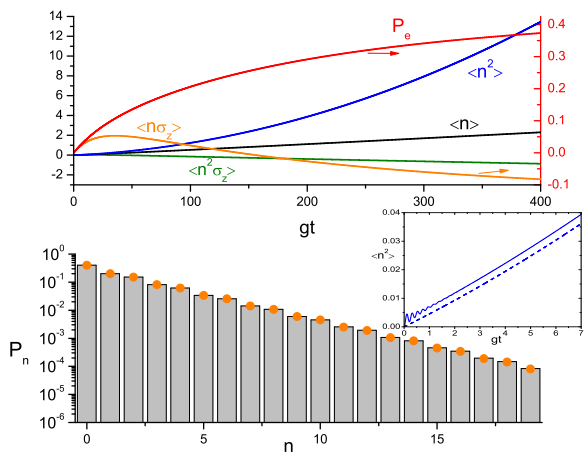}
\end{center}
\caption{Same as figure \protect\ref{fig1} in the dispersive regime, $\Omega
=\protect\omega -20g$.}
\label{fig2}
\end{figure}

\begin{figure}[th]
\begin{center}
\includegraphics[width=.49\textwidth]{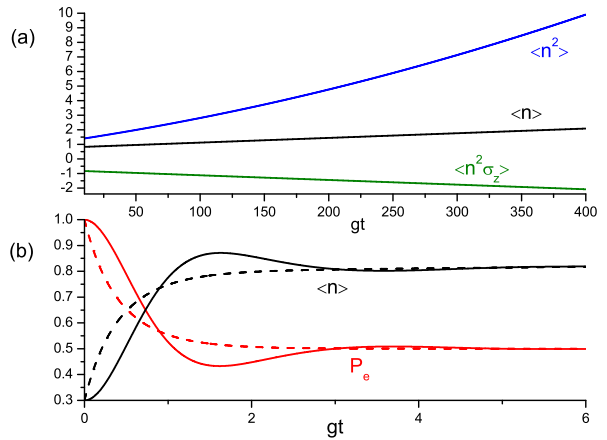}
\end{center}
\caption{Behavior of $\left\langle n\right\rangle $, $\left\langle
n^{2}\right\rangle $, $\left\langle n^{2}\protect\sigma _{z}\right\rangle $
and $P_{e}$ in \textbf{(a)} the asymptotic regime and \textbf{(b) }during
the transient regime for small times. The initial state is $\protect\rho %
_{therm}\otimes |e\rangle \langle e|$ with average photon number $%
\left\langle n(0)\right\rangle =0.3$.}
\label{fig3}
\end{figure}

In the figures \ref{fig1}--\ref{fig3} we compare the exact dynamics
resultant from the original differential equations (\ref{a1})--(\ref{a3}) to
the effective dynamics governed by the simplified equations (\ref{d1})--(\ref%
{d2}) for the parameters $\omega =1$, $g=4\times 10^{-2}$, $\gamma _{a}=2g$
and $\gamma _{c}=0$. In the figure \ref{fig1} (\ref{fig2}) we consider the
resonant regime $\Omega =\omega $ (dispersive regime $\Omega =\omega -20g$)
for the initial zero-excitation state $|g,0\rangle $. We plot the dynamical
behavior of observables $\left\langle n\right\rangle $, $\left\langle
n^{2}\right\rangle $, $\left\langle n^{2}\sigma _{z}\right\rangle $, $%
\left\langle n\sigma _{z}\right\rangle $ and $P_{e}$ calculated from the
original differential equations (\ref{a1})--(\ref{a3}). Within the thickness
of the lines these curves are indistinguishable from the graphs resulting
out of the effective equations (\ref{d1})--(\ref{d2}). To exemplify the
difference between the original and effective differential equations, we
show the zoom for the behavior of $\left\langle n^{2}\right\rangle $ at
initial times: the solid line depicts the exact dynamics, and the dashed
line -- the effective one. The observed discrepancies are quite small and
appear because $c_{n,m}^{\prime }$ does not becomes zero instantly as was
assumed in our analysis; nevertheless, these minor differences are
irrelevant regarding the asymptotic behavior. In the figures we also show
the photon number distributions calculated numerically at the time $gt=300$
according to the original differential equations (bars) and the effective
ones (dots). Once again, the agreement is excellent.

In the figure \ref{fig3} we consider the initial state $\rho (0)=\rho
_{therm}\otimes |e\rangle \langle e|$, where $\rho _{therm}$ is the thermal
state of the Electromagnetic field whose photon number distribution is $%
p_{n}=\bar{n}^{n}/(\bar{n}+1)^{n+1}$, where $\bar{n}$ is the average photon
number. We set $\Omega =\omega $, $\bar{n}=0.3$ and show the asymptotic
behavior of $\left\langle n\right\rangle $, $\left\langle n^{2}\right\rangle
$ and $\left\langle n^{2}\sigma _{z}\right\rangle $ obtained from the
original differential equations (figure \ref{fig3}a) and the zoom of $%
\left\langle n\right\rangle $ and $P_{e}$ for initial times (figure \ref%
{fig3}b, where the dashed lines represent the solutions of the effective
differential equations). We see that asymptotically the behavior agrees with
equations (\ref{nn})--(\ref{x4}), although the transient dynamics cannot be
reliably described by the equations (\ref{d1})--(\ref{d2}).

Regarding the practical observation of the asymptotic linear photon growth inside the cavity due
to decoherence, it seems quite unlikely in current cavity or
circuit QED implementations because the photons (and atomic excitations)
would be lost due to radiative and nonradiative relaxation processes. As
example, let us consider the current state of the art circuit QED
implementations. The typical parameters are \cite{Fedor}: $\omega \sim \Omega
\sim 8\,$\textrm{GHz} and $g\sim 0.3\,$\textrm{GHz}, while the dephasing
rate is of the order of $\gamma _{a}\sim 1\,\mathrm{MHz}$, although it can
be made large at will (usually one desires to decrease $\gamma $ and not to
increase it). Considering a high value for the total dephasing rate $\gamma
\sim 1\mathrm{GHz}$ the resulting asymptotic photon growth rate due to
dephasing would be $\sim 1\,\mathrm{MHz}.$ This value is of the same order
of magnitude as the cavity relaxation rate for a rather high cavity quality
factor $Q\sim 10^{4}$, so $\left\langle n(t)\right\rangle $ would saturate
at some (small) value instead of showing an asymptotic growth, as calculated
explicitly in \cite{CAMOP,2atoms,amendart} for standard quantum optical master
equation. Some photons escape to the outside world via radiative dissipation
channel so they could be ultimately detected outside the cavity, but in this
case different models predict different photon emission rates depending on
assumptions made about the reservoir \cite{Libe,W3} (in particular
whether it is Markovian or not).

Recently a more sophisticated microscopic
model was developed for deducing the master equation in the presence of the anti-rotating term, valid in a specific regime of parameters \cite{Bla}. According to that model, the phenomenon of dephasing-induced generation of photons is greatly exaggerated by the Lindblad-type master equation (\ref{111}),
and instead of the linear asymptotic growth the average photon number saturates at
some value that strongly depends on the reservoir spectral density \cite{Bla}. Nevertheless, the very phenomenon of photon generation due to decoherence persists and our formulae provide
the upper bound for the photon generation rate. From the qualitative viewpoint, in realistic (lossy) cavity QED architectures this phenomenon would lead to a parameter-dependent heating of the system slightly above the thermal equilibrium values \cite{2atoms,amendart}, depending on the atom-field detuning, coupling strength and the dephasing rate.

In summary, we obtained simplified differential equations describing the
process of photon generation (from vacuum or any other state) due to the
combined action of the anti-rotating term and the standard Lindbladian dephasing in Markovian
cavity QED, whose validity was confirmed by extensive numerical simulations.
From these equations we deduced analytical formulae describing the overall
behavior of MENDART for arbitrary initial state, demonstrating that
asymptotically the mean photon number $\left\langle n\right\rangle $
increases linearly with time at the rate $2\gamma g^{2}(\omega ^{2}+\Omega
^{2}+\gamma ^{2})^{-1}$, $\left\langle n^{2}\right\rangle $ grows
quadratically with time and the atomic excitation probability attains a constant value.

\begin{acknowledgments}
The author acknowledges a partial support of Decanato de Pesquisa e P\'{o}%
s-Gradua\c{c}\~{a}o (Universidade de Bras\'{\i}lia, Brazil).
\end{acknowledgments}

\end{document}